\documentclass[twocolumn,prX]{revtex4}
\usepackage{epsfig, graphicx}
\usepackage{float}
\usepackage{amsmath}
\usepackage{latexsym}
\usepackage{epsfig, graphicx} 
\usepackage{graphics}

\newcommand{\be}{\begin{equation}}
\newcommand{\bel}[1]{\begin{equation}\label{#1}}
\newcommand{\ee}{\end{equation}}

\newcommand{\ds}{\displaystyle}
\newcommand{\beq}{\begin{eqnarray}}
\newcommand{\eeq}{\end{eqnarray}}
\newcommand{\beqq}{\begin{eqnarray*}}
\newcommand{\eeqq}{\end{eqnarray*}}
\newcommand{\p}{\partial}

\begin{document}
\title{Quantifying intermittent transport in cell cytoplasm}
\author{Thibault Lagache}
\address {Department of Biology, Ecole Normale Sup\'erieure, 46 rue
d'Ulm 75005 Paris, France}
\author{David Holcman}
\address{Department
of Mathematics, Weizmann Institute of Science, Rehovot 76100,
Israel} \address {Department of Biology, Ecole Normale Sup\'erieure, 46 rue d'Ulm 75005 Paris,
France}
\begin{abstract}
Active cellular transport is a fundamental mechanism for protein and
vesicle delivery, cell cycle and molecular degradation. Viruses can
hijack the transport system and use it to reach the nucleus. Most
transport processes consist of intermittent dynamics, where the
motion of a particle, such as a virus, alternates between pure
Brownian and directed movement along microtubules. In this
communication, we estimate the mean time for particle to attach to a
microtubule network. This computation leads to a coarse grained
equation of the intermittent motion in radial and cylindrical
geometries. Finally, by using the degradation activity inside the
cytoplasm, we obtain refined asymptotic estimations for the
probability and the mean time a virus reaches a small nuclear pore.
\end{abstract}


\maketitle

{\noindent \textit{Introduction.}}
Cell transport, which may involve vesicles or proteins is
essential for cellular function and homeostasis. In general free
diffusion in the cell cytoplasm is not efficient and many
particles such as large viruses cannot pass the crowded cytoplasm
\cite{sodeik} without hijacking the complex cellular transport
machinery and use molecular motors, such as dyneins, to travel
along microtubules (MTs) toward the nucleus.  Both vesicular and
viral motions alternate intermittently between periods of free
diffusion and directed motion along MTs \cite{Greber}. Such viral
trajectories have been recently monitored by using new imaging
techniques \textit{in vivo} \cite{charneau,Seisenberger}.

The switch nature of the motion, imposes a complex behavior of the
particle trajectories which depends on the number and distribution
of MTs, the rate of binding and unbinding and the diffusion
constant of the free particle. Some physical properties, such as
the mean velocity of trajectories has been obtained for the motion
in domain made of parallel strips, in which a random particle has
a deterministic motion on the stripes and pure diffusion outside
\cite{ajdari}. In case of a population of motors, at equilibrium
between free diffusion and bound on MTs, the motor distribution has
been studied in cylindrical and radial geometries in
\cite{nedelec,lipo1}; the authors estimate the forward binding rate
using Brownian simulations in \cite{lipo1} and experimentally in
\cite{nedelec}.

We consider here a particle $\mathbf{x}(t)$, which can be
described using the stochastic rule:
 \beq
 d\mathbf{x}= \left\{\begin{array}{l} \sqrt
{2D}
d\mathbf{w} \quad\mbox{ for }\quad \mathbf{x}\left(t\right) \quad\mbox{ free }\\ \\
{\bf V} \quad\mbox{ for }\quad \mathbf{x}\left(t\right)
\quad\mbox{ bound }
\end{array}\right.\label{eq1}
\eeq where $\bf w$ is a standard Brownian motion, $D$ the
diffusion constant and $\bf V$ the velocity of the directed motion
along MTs.

In this communication, we compute the mean first passage time of a
single particle to a population of MTs. We thus provide an
analytical expression of the forward binding rate of a motor to
MTs in both radial and cylindrical geometries. Using the
analytical expression of the forward binding rate, we propose a
coarse-grained description of a switch dynamical motion of a
particle, which can either be a virus, a vesicle or a molecular
motor. This description, which is the main result of our paper, is
a fundamental step to estimate the probability and the mean time
to arrive at a small target. Moreover, using this description, we
obtain the steady state distribution of virus in MTs network
without resorting to the assumption of a two-state model
\cite{nedelec,lipo2}.

We thus compute an effective steady state drift $\mathbf{b(x)}$ such
that the particle motion (\ref{eq1}) can be coarse-grained by the
stochastic equation: \beq d\mathbf{x} = \mathbf{b(x)}dt+\sqrt{2D}d
\bf w.
 \label{langevin}
 \eeq
 Using results derived in \cite{david}, equation (\ref{langevin}) and the degradation
activity in the cytoplasm due to protease or lysosome,
we obtain asymptotic estimates of the probability and the mean
time for a virus to reach a nuclear pore. The problem of finding a
small target is ubiquitous in cellular biology and recent
theoretical studies \cite{PNAS,klafter} suggest that the
geometrical organization of the medium play a fundamental role in
this search process.
\par
{\noindent \textit{Mathematical Modeling.}}
We represent the cell cytoplasm as a bounded domain $\Omega$, whose
boundary $\partial \Omega $ consists of the external membrane
$\partial \Omega _{ext}$ and the nuclear envelope, both of which
form a reflecting boundary $\partial N_r $ for the trajectories of
(\ref{langevin}), except for small nuclear pores $\partial N_a $,
where they are absorbed. The ratio of boundary surface areas
satisfies $ \varepsilon=\frac{|\p N_a|}{|\p \Omega|} \ll 1.$ We
model the virus degradation activity in the cell cytoplasm as a
steady state killing rate $k(\bf{x})$ for the trajectories of
(\ref{langevin}), so the survival probability density function
(SPDF) is the solution of the Fokker-Planck equation \cite{HMS}
\[
\begin{array}{c}
 \ds{\frac{\partial p}{\partial t}}=D\Delta p-\nabla \cdot \mathbf{b}p-kp \\
 p\left( {\mathbf{x},0} \right)=p_i \left(\mathbf{x} \right) \\
 \end{array}
\] with the boundary conditions:
\beq p(\mathbf{x},t) = 0 \hbox{ on } \p N_a \hbox{ and }
\mathbf{J}(\mathbf{x},t).\mathbf{n}_{\mathbf{x}} = 0 \hbox{ on }
\p N_r \cup \p \Omega_{ext} \label{boundary-c2} \eeq where
$\mathbf{n}_{\mathbf{x}}$ denotes the normal derivative at a
boundary point $\mathbf{x}$. The flux density vector
$\mathbf{J}(\mathbf{x},t)$ is defined as
 \beq
J(\mathbf{x},t)=-D \nabla p(\mathbf{x},t)
+\mathbf{b(x)}p(\mathbf{x},t).\label{Ji}
 \eeq
The probability $P_N $ and the mean time $\tau _N $ that a
trajectory of (\ref{langevin}) reaches $\partial N_a $ are given
by the small hole theory for two-dimensional domains $\Omega$ and
drifts $b\left( x \right)=-\nabla \Phi \left( x \right)$, as
\cite{david}
\beq \left\{\begin{array}{l}
P_N=
\frac{\frac{1}{|\partial \Omega|}\int_{\partial
\Omega}e^{-\frac{\Phi(\mathbf{x})}{D}}dS_{\mathbf{x}}}
{\frac{ln\left(\frac{1}{\epsilon}\right)}{D\pi}\int_{\Omega}e^{-\frac{\Phi(\mathbf{x})}{D}}k(\mathbf{x})d\mathbf{x}+\frac{1}{|\partial \Omega|}\int_{\partial \Omega}e^{-\frac{\Phi(\mathbf{x})}{D}}dS_{\mathbf{x}}} , \label{dim3} \\ \\
\tau_N =
\frac{\frac{ln\left(\frac{1}{\epsilon}\right)}{D\pi}\int_{\Omega}e^{-\frac{\Phi(\mathbf{x})}{D}}d\mathbf{x}}
{\frac{ln\left(\frac{1}{\epsilon}\right)}{D\pi}\int_{\Omega}e^{-\frac{\Phi(\mathbf{x})}{D}}k(\mathbf{x})d\mathbf{x}+\frac{1}{|\partial
\Omega|}\int_{\partial
\Omega}e^{-\frac{\Phi(\mathbf{x})}{D}}dS_{\mathbf{x}}},\label{2D}
\end{array}\right.
\label{tau-ff} \eeq Hereafter we derive explicitly the steady
state drift $\mathbf{b(x)}$ as a function of some geometrical and
dynamical parameters of the cell (number of MTs) and the virus
(binding and unbinding rates, the mean velocity $\mathbf{V}$ of
the directed motion and the diffusion constant $D$). \par
\begin{figure}
\includegraphics[width=6cm]{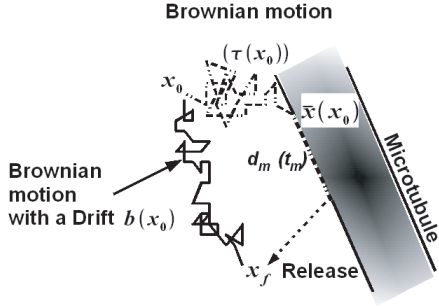}
\caption{\small{ The fundamental step is represented with a dotted
line; a virus starts at a position $\mathbf{x_0}$, diffuses
freely, binds to a MT over a distance $d_m$ and is then released
at a final position $\mathbf{x_f}$. The solid line represents a
trajectory generated by the steady state equation (\ref{langevin}). In the
parenthesis, we point out the mean times for each portion of
trajectories.}}\label{scheme_step}
\end{figure}

{\noindent \textit{General Methodology.}}
To derive an expression for $\mathbf{b(x)}$, we consider the motion of a virus between the
moment it enters the cell at the outer membrane and the moment it
reaches the absorbing boundary $\partial N_a $. Its motion alternates
between free diffusion, for a random time $\tau$, until it hits a
MT and binds. It continues in a directed motion along the MT for a
mean time $t_m $ and a mean distance $d_m =\left\| V \right\|t_m$,
until it is released and resumes free diffusion. The steady state
drift is chosen to be constant for a sufficiently small step,
such that the mean time $\tau+t_m $ to the first release at a point $\mathbf{x_f}$
is the same as that predicted by (\ref{langevin}) (see
FIG. \ref{scheme_step}). This approach leads to explicit expressions for the steady
state drift for two-dimensional radial and cylindrical geometries.
\par
\begin{figure}
\includegraphics[width=5.5cm]{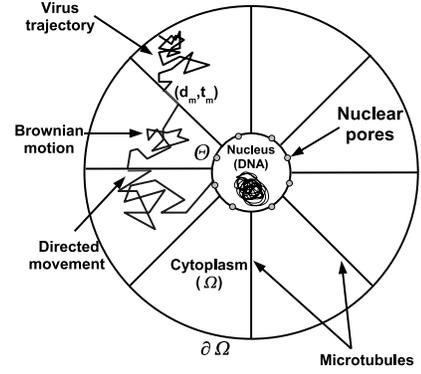}
\caption{\small{Two dimensional radial cell with radially
equidistributed MTs. We show a virus trajectory alternating
between bound and diffusive periods in cytoplasm.}}
\label{cellule_radial}
\end{figure}
\begin{figure}
\includegraphics[width=5.5cm]{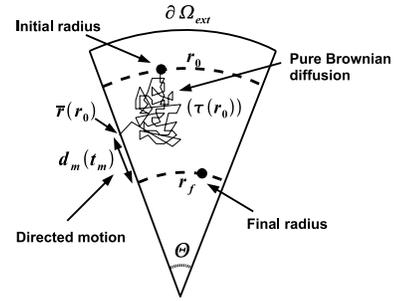}
\caption{\small{ A fundamental step in $\tilde{\Omega}$. The virus
starts at a radius $r_0$, with an angle uniformly distributed in
$[0;\Theta]$, it diffuses freely during a time $\tau(r_0)$ until
it binds to a MT at a {mean} radius $\bar{r}(r_0)$; it has then a
directed motion over a distance $d_m=||\mathbf{V}||t_m$ before
being released randomly at a final radius $r_f$. Mean times of
each piece of the fundamental step are written inside
parenthesis.}}\label{iteration}
\end{figure}

{\noindent \textit{The steady state drift for a two-dimensional
radial cell.}}
We consider a two-dimensional cell cytoplasm which is an annulus
$\Omega$ of outer radius $R$ and inner radius $\delta$ (nuclear
surface) with $N$ MTs radially uniformly distributed. They
irradiate from the nucleus to the external membrane and the angle
between two neighboring ones is $\Theta=\frac{2\pi}{N}$. The
two-dimensional approximation applies for culture cells which are
flat \cite{dinh} due to the adhesion to the substrate. In that
case the thickness can be neglected in the computation. Before
reaching a small nuclear pore, a virus has an intermittent
dynamics, alternating between diffusing and bound periods (see
FIG. \ref{cellule_radial}).
Because the MTs are uniformly distributed, we consider the
fundamental domain $\tilde{\Omega}$ defined as the two-dimensional
slice of angle $\Theta$ between two neighboring ones. In
$\tilde{\Omega}$, the fundamental step described above is as
follows: the virus starts at a radius $r_0$ with an angle
uniformly distributed in $[0;\Theta]$, it binds to a MT at a time
$\tau(r_0)$ and at a radius $\bar{r}(r_0)$. On the MT, it has a
radially directed movement towards the nucleus during a {mean}
time $t_m$ and over a distance $d_m=||\mathbf{V}||t_m$. Finally,
the virus is released with a $\Theta$-uniformly distributed angle
at a final radius $r_f=\bar{r}(r_0)-||\mathbf{V}||t_m$ (see FIG.
\ref{iteration}).
In most eukaryotic cell large asters,  there are from $600$ to
$1000$ MTs \cite{nedelec}. We can estimate the average number $N$ of
MTs per cell cross section as follow: for a cell thickness $h\approx
9\mu m$, \cite{nedelec}, an interaction range $\gamma \approx 50nm$
between the MTs and the molecular motors \cite{coy}, and for the AAV
diameter $d=30nm$ \cite{Seisenberger}, we obtain for a radial MT
organization in a thin cylindrical cell, that the range of $N$ is
between $\frac{600(2\gamma +d)}{h}$ and $\frac{1000(2\gamma
+d)}{h}$, that is  $9$ to $15$. We are thus in a regime where
$\Theta <<1$. For $r_0<R$, by neglecting the reflecting external
boundary at $r=R$, $\tilde{\Omega}$ becomes an open wedge and thus
using the standard methods from
\cite{redner,lagache}, we obtain \beq \tau(r_0)\approx r_0^2
\frac{\Theta^2}{12 D} \hbox{ and } \bar{r}(r_0) \approx r_0(1+
\frac{\Theta^2}{12}). \label{tib} \eeq In radial geometry,
$\mathbf{b(x)}=b(r)\frac{\mathbf{r}}{||\mathbf{r}||}$ and the MFPT
$u(r_0)$ of a virus starting at $r_0$ and ending at position
$r_f$, described by equation (\ref{langevin}) satisfies
\cite{Schuss}:
\begin{eqnarray}
D\Delta u-b(r_0)\nabla u = -1 \label{eqou}\\
\frac{d u}{dr}(R)=0 \hbox{ and } u(r_f)=0, \nonumber
\end{eqnarray}
where we approximated $b(r)$ by $b(r_0)$. The solution of equation
(\ref{eqou}) is
\begin{eqnarray}
u(r_0)=\int_{r_f}^{r_0}\left(\int_{v}^{R}\frac{ue^{-\frac{b(r_0)}{D}(u-v)}}{Dv}
du \right) dv.
\end{eqnarray}
For $D<<1$, using the Laplace method,
 \beq
\int_{v}^{R}\frac{ue^{-\frac{b(r_0)}{D}(u-v)}}{Dv} du \approx
\frac{1}{b(r_0)}. \eeq
Thus, in first approximation, $ u(r_0)\approx
\frac{r_0-r_f}{b(r_0)}$. To obtain the value $b(r_0)$,
we equal the MFPT $u(r_0)$ from $r_0$ to $r_f$ computed from
equation (\ref{langevin}) with the one obtained from an
intermittent dynamic: $\tau(r_0)+t_m$. Consequently, we get:
\beq \label{eqfi}
b(r_0)=\frac{r_0-r_f}{\tau(r_0)+t_m}=\frac{d_m-r_0\frac{\Theta^2}{12}}{t_m+r_0^2
\frac{\Theta^2}{12D}}.
\eeq
\par
\begin{figure}
\includegraphics[width=6cm]{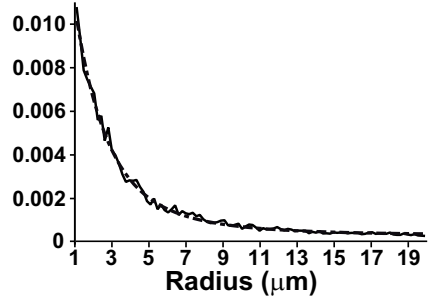}
\caption{\small{ Steady state distributions. Dashed line: virus
distribution (\ref{sol}) with the effective drift $b(r)$
(\ref{eqfi}); solid line: Empirical steady state distribution
obtained by running $10,000$ intermittent Brownian trajectories}.The
cell radius is $R=20 \mu m$ and $\Theta=\frac{\pi}{6}$.
}\label{FIGURE2}
\end{figure}

{\noindent \textit{Tests against Brownian simulations.}}
We impose reflecting boundaries at the external membrane $r=R$ and
we tested the theoretical steady state distribution against the
one obtained by running empirical intermittent Brownian
trajectories in the pie wedge domain. For a potential field, the
steady state distribution satisfies $D\Delta p -
\nabla[\mathbf{b}p]=0 \hbox{ in } \Omega$ with  reflecting
boundary condition
$\mathbf{J}(\mathbf{x},t).\mathbf{n}_{\mathbf{x}} = 0 \hbox{ on }
\partial \Omega.$ The distribution p in a two-dimensional radial geometry is:
\beq \label{sol}
p(r)=\frac{e^{-\frac{\Phi(r)}{D}}}{\int_{0}^{R}e^{-\frac{\Phi(r}{D}}2\pi
r dr}, \eeq which should be compared to the distribution of
\cite{nedelec}. The potential $\Phi$ of $b=-\nabla \Phi$ is
obtained by integrating equation (\ref{eqfi}) with respect to $r$,
\beq \Phi(r)=\frac{d_m\sqrt{12 D t_m}}{t_m\Theta}
\arctan\left(\frac{\Theta r}{\sqrt{12 D t_m}}\right)\\
-\frac{D}{2}
\ln\left(12 D t_m+r^2\Theta^2\right) \nonumber
\eeq
%
In FIG. \ref{FIGURE2}, we plotted the steady state distribution
given in (\ref{sol}) against the distribution obtained by the
intermittent empirical equation (\ref{eq1}). The parameters are
chosen such that the viruses move towards the nucleus (observed
\textit{in vitro}, loaded dynein moves during $1s$ over a distance
of $0.7\mu m$
\cite{schroer}), we thus take $t_m=1s$ and $d_m=0.7 \mu m$;
furthermore, the diffusion constant is $D=1.3\mu m^2 s^{-1}$ as
observed for the Associated-Adeno-Virus
\cite{Seisenberger}. The nice agreement of both curves, 
which is the central result of this communication, confirms that
our coarse grained method accounts well for the switch system
(\ref{eq1}). 
\par
{\noindent \textit{Computation of $P_N$ and $\tau_N$}.}
We derive now asymptotic expressions in the small diffusion limit
$D<<1$,  for the probability $P_N$  and the mean time $\tau_N$ a
virus arrives at a small nuclear pore. We apply Laplace's method in
formulas (\ref{2D}) for a radial geometry. When the degradation rate
$k(r)$ is taken constant, equal to $k_0$ in the neighborhood of the
nucleus $r=\delta$ and when $12d_m > r \Theta^2$, $\hbox{ }b(r)>0$
so that $\Phi$ reaches its minimum at $r=\delta$, we get
\begin{equation*} P_N= \ds{\frac{b(\delta)}
{ln\left(\frac{1}{\epsilon}\right)2\delta k_0+ b(\delta)}} \hbox{
and } \tau_N = \ds
{\frac{ln\left(\frac{1}{\epsilon}\right)2\delta}
{ln\left(\frac{1}{\epsilon}\right)2\delta
k_0+b(\delta)}}.\label{tau-ffp} \end{equation*}
%
A Taylor expansion for $\Theta\ll 1$ gives that
\begin{eqnarray}
P_N&\approx&\frac{d_m}{d_m+K} \left(1-\frac{K\delta\left(d_m\delta+Dt_m\right)}{12Dt_m d_m\left(d_m+K\right)}\Theta^2\right)\label{1}\\
\tau_N&\approx&\frac{K}{k\left(d_m+K\right)}\left(1+\frac{\delta\left(d_m\delta+Dt_m\right)}{12Dt_m\left(d_m+K\right)}\Theta^2\right)\label{2}
\end{eqnarray}
where $K=2k_0\delta t_m \ln\left(\frac{1}{\epsilon}\right)$.
\par
%
We can now propose the following predictions: because nuclear pores
occupy a fraction $\epsilon=2\%$ \cite{maul} of the nucleus surface
(radius $\delta=8\mu m$) and the measured degradation rate for
plasmids \cite{Lechardeur} is $k=\frac{1}{3600}s^{-1}$, we obtain
from formula \ref{1}-\ref{2} that \beq P_N \approx 94.3\%\hbox{,
}\tau_N \approx 205s.
\eeq
We conclude that the infection efficiency is very high, while the
mean time to reach a nuclear pore is of the order of 3 minutes. It
is interesting to compare this time with the 15 minutes reported in
\cite{Seisenberger}, which accounts for all the viral infection steps from the entry to
the final nuclear import. This difference between the two times
indicates that the phase where the virus is inside an early endosome
(EE) may last 10 minutes. Indeed, the endosomal phase ends once the
EE has matured into a late endosome (LE) \cite{engelhardt}, which
lasts approximately $10$ minutes \cite{zerial}. To finish, we shall
note that a free diffusing virus would reach a nuclear pore in about
15 minutes \cite{PNAS}.
\par
\begin{figure}
\includegraphics[width=5.5cm]{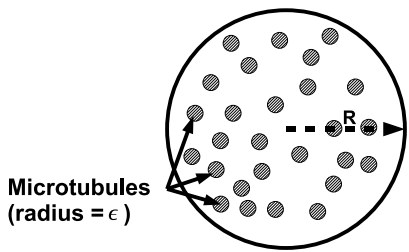}
\caption{\small{Dendrite cross-section. The $N$ MTs are thin
cylinders uniformly distributed inside the
 dendrite.}}\label{dendrite}
\end{figure}

{\noindent \textit{The cylindrical geometry.}}
Many transports mechanisms such as viral (herpes virus
\cite{smith}) and vesicular occur in long axons or dendrites,
which can be approximated as thin cylinders (radius $R$ and length
$L$). To derive a quantitative analysis of viral infection in that
case, we follow the method described above and compute the steady
state drift that accounts for the directed motion along MTs. We
model the $N$ MTs parallel to the dendrite {principal} axis as
cylinders (radius $\epsilon<<R$, Length $L$). The cross-section
$\Omega$ of the dendrite is shown in FIG. \ref{dendrite}.
 Due to the cylindrical symmetry, for any position
$\mathbf{x}$, the steady state drift $\mathbf{b(x)}$ is equal to $B
\mathbf{z}$ where $B$ is a constant and $\mathbf{z}$ the principal
axis unit vector along the dendrite. In a small diffusion
approximation, the leading order term of $B$   is equal to the
effective velocity \cite{ajdari,lipo1}: $B=\frac{d_m}{t_m+\tau},$
where $t_m$ is the mean time the virus binds to a MT,
$d_m=||\mathbf{V}||t_m$ the mean distance of a run and $\tau$ the
MFPT to a MT. To derive an expression for $\tau$, we consider the
cross-section  $\Omega$ and impose reflecting boundary condition at
the external membrane of the dendrite ($r=R$) and absorbing ones at
the MTs surfaces. In long time approximation, for a MTs radius
$\epsilon<<1$, $\tau$ is asymptotically equal to $\frac{1}{\lambda
D}$ where $\lambda$ is the first eigenvalue of the Laplace operator
in $\Omega$ with the boundaries conditions described above
(\cite{Schuss} p.175). The leading order term of $\lambda$ as a
function of $\epsilon$ is \cite{osawa} $\lambda=\frac{2\pi
N}{|\Omega|ln\left(\frac{1}{\epsilon}\right)},$ where $|\Omega|=\pi
R^2$. Thus, the MFPT to a MT is $
\tau=\frac{1}{\lambda
D}=\frac{R^2ln\left(\frac{1}{\epsilon}\right)}{2 ND},$ and the
steady state drift amplitude $B$ is given by \beq
B=\frac{d_m}{t_m+\tau}=\frac{2 N D d_m}{2 N D t_m +
R^2ln\left(\frac{1}{\epsilon}\right)}. \eeq We conclude that in the
limit $t_m \ll \tau$, the effective velocity is proportional to the
number of MTs: $B\approx N \frac{2D d_m}{R^2
ln\left(\frac{1}{\epsilon}\right)}$, as already observed in
\cite{nedelec}.

{\noindent \textit{Conclusion.}}
Intermittent dynamics with alternative periods of free diffusion and
directed motion along MTs characterizes many cellular transports. We
have developed a model to estimate a steady state drift such that
the intermittent dynamic can be described by an over-damped limit of
the Langevin equation. Our method gives explicit results in
two-dimensional {radial} cell and in a cylindrical dendrite or axon.
The steady state description of the movement enables us to estimate
the probability a virus reaches alive a small nuclear pore and its
mean time. Because viruses are very efficient DNA carriers,
understanding and quantifying their movement in the cell cytoplasm
would be very helpful for designing synthetic vectors \cite{Zuber}.
In a future work, it would be interesting to derive steady state
drifts for three dimensional geometries.

{\bf Acknowledgments:} D. H. research is partially supported by
the program ERC-starting Grant.

\end{document}